\documentclass[prd,aps,lengthcheck,twocolumn,nofootinbib,notitlepage,floatfix,superscriptaddress,preprintnumbers]{revtex4-2}
\usepackage{graphicx} 
\usepackage{amsmath,amssymb}
\usepackage{lipsum}
\usepackage{bm}
\usepackage{braket}
\usepackage{color,xcolor}
\usepackage{comment}
\usepackage[colorlinks=true,allcolors=blue,breaklinks]{hyperref}

\begin{document}
\title{Thermodynamics of fermionic excitations in heavy-quark QCD}
\author{Kei Tohme}
\email{tohme@ruby.scphys.kyoto-u.ac.jp}
\affiliation{Division of Physics and Astronomy, Graduate School of Science, Kyoto University,
Kitashirakawaoiwake, Sakyo, Kyoto 606-8502, Japan}
\author{Takahiro M. Doi}
\email{doi.takahiro.5d@kyoto-u.ac.jp}
\affiliation{Division of Physics and Astronomy, Graduate School of Science, Kyoto University,
Kitashirakawaoiwake, Sakyo, Kyoto 606-8502, Japan}
\author{Masakiyo Kitazawa}
\email{kitazawa@yukawa.kyoto-u.ac.jp}
\affiliation{Yukawa Institute for Theoretical Physics, Kyoto University, Kyoto, 606-8502, Japan}
\affiliation{J-PARC Branch, KEK Theory Center, Institute of Particle and Nuclear Studies, KEK, 319-1106, Japan}
\author{Krzysztof Redlich}
\email{krzysztof.redlich@uwr.edu.pl}
\affiliation{Institute of Theoretical Physics, University of Wroc\l{}aw,  Wroc\l{}aw PL-50204, Poland}
\affiliation{Polish Academy of Sciences PAN, 
Wroc\l{}aw PL-50449, Poland}
\author{Chihiro Sasaki}
\email{chihiro.sasaki@uwr.edu.pl}
\affiliation{Institute of Theoretical Physics, University of Wroc\l{}aw, Wroc\l{}aw PL-50204, Poland}
\affiliation{International Institute for Sustainability with Knotted Chiral Meta Matter (WPI-SKCM$^2$), Hiroshima University, Hiroshima 739-8526, Japan}

\date{\today}

\preprint{YITP-25-121, J-PARC-TH-0321}

\begin{abstract}
We investigate the thermodynamic properties of fermionic excitations in heavy-quark QCD on the lattice with Wilson fermions.
The grand potential is calculated analytically in the hopping parameter expansion (HPE) on the basis of the cumulant expansion.
Using the grand potential, we compute the quark number susceptibilities and their ratios up to next-to-leading order in the HPE. The ratio of fourth- to second-order susceptibilities is shown to be unity (nine) in the deconfined (confined) phase at the leading order.
Excitation properties of baryonic and quark modes in each phase are also investigated utilizing the Boltzmann statistics.
We obtain an analytic formula for the quark excitation energy in the deconfined phase, while that for baryonic excitations in the confined phase is decomposed into flavor multiplets.
\end{abstract}

\maketitle

\section{Introduction}
\label{sec:intro}

Thermal properties of heavy-mass excitations are governed by Boltzmann statistics, which results in their simple description in thermodynamics. 
This applies to baryonic excitations in Quantum Chromodynamics (QCD) below the pseudo-critical temperature $T_{\rm pc}$ at zero baryon density. 
An interesting outcome from it is that all even-order baryon-number susceptibilities $\chi^{\rm B}_n$ ($n\in2\mathbb{N}$) are equivalent in magnitude~\cite{Ejiri:2005wq,Asakawa:2015ybt}. In fact, lattice-QCD numerical simulations show that the ratio $\chi^{\rm B}_4/\chi^{\rm B}_2$ is consistent with unity below $T_{\rm pc}$~\cite{Ejiri:2005wq,Allton:2005gk,Borsanyi:2013hza,Bazavov:2013dta,Bazavov:2020bjn,Borsanyi:2023wno}. The lattice results also show that the ratio suddenly drops from unity above $T_{\rm pc}$, indicating the violation of this picture, i.e., the onset of the quark deconfinement. Similar behaviors have been confirmed in various mixed susceptibilities including strange and charm degrees of freedom~\cite{Bazavov:2013dta,Bellwied:2013cta,Bazavov:2023xzm,Kaczmarek:2025dqt}.

In exploring thermodynamic properties of QCD, varying the quark masses from the physical values provides a valuable theoretical framework for gaining alternative insights into the real world~\cite{Pisarski:1983ms,Philipsen:2021qji}. 
In the heavy-mass limit, QCD reduces to SU(3) Yang-Mills (YM) theory. The existence of a first-order phase transition separating the confined and deconfined phases at nonzero temperature $T$ in this theory suggests that the excitations carrying quark charges change suddenly from the baryonic ones to quarks at $T=T_{\rm c}$ in this limit. Besides, as the Boltzmann statistics should be well applicable to quark excitations in the deconfined phase owing to heavy quark masses, the even-order {\it quark}-number susceptibilities $\chi_n$ should be equivalent, where $\chi_n$ are related to baryon-number susceptibilities as $\chi_n=3^n\chi^{\rm B}_n$. On the other hand, for $T<T_{\rm c}$ the quark charges will be carried by baryonic states and all baryon-number susceptibilities are equivalent as in the case of QCD with physical quark masses. Therefore, the higher order susceptibilities are expected to change discontinuously at $T=T_{\rm c}$ between the two asymptotic values.

In the present study, we examine the validity of this picture in the heavy-mass limit of QCD, which we refer to as heavy-quark QCD in what follows, based on the lattice-QCD formulation with Wilson fermions. For large quark masses, this formalism allows one to adopt the hopping-parameter expansion (HPE), which corresponds to an expansion of the quark action by the inverse quark masses. With the aid of the cumulant expansion, we reorganize the HPE of the grand potential so that it is classified into terms carrying different chemical-potential dependence, and calculate several leading terms analytically. As a result, our grand potential consists of the expectation values of the gluonic operators in the heavy-mass limit, which are numerically calculable in the Monte Carlo simulations of ${\rm SU}(3)$ YM theory. The obtained result is quite convenient for interpreting their physical meaning and calculating susceptibilities. We notice that no assumptions are imposed on the gluonic part in this manipulation.  

Using the analytic expression of the grand potential, we reveal two important features of the fermionic excitations in heavy-quark QCD on the lattice. First, we calculate the HPE of the ratios of quark-number susceptibilities, $\chi_{n+n'}/\chi_n$, up to next-to-leading (NLO) order for $n'\in2\mathbb{N}$. We show that at the leading-order (LO) the ratio is exactly one and $3^{n'}$ above and below $T_{\rm c}$, respectively, as anticipated from the aforementioned argument. We show that this result is a direct consequence of the spontaneous symmetry breaking of the $\mathbb{Z}_3$ symmetry at $T=T_{\rm c}$ in SU(3) YM theory. It will also be shown that the order of the NLO terms is significantly high due to cancellations of the expansion terms in the ratio as will be discussed in Sec.~\ref{sec:NLO}. 

Second, we explore properties of fermionic excitations in the confined and deconfined phases in more detail using the analytic formula of the grand potential and Boltzmann statistics. We show that the excitation energy of quarks in the deconfined phase is given by the sum of the pole mass and the free energy of a single quark, the latter of which is given by the Polyakov loop. We also relate the baryonic excitations in the confined phase with the expectation values of the loop operators having three temporal windings, and classify them into flavor multiplets. We show that the excitation reduces to a simple picture in the strong coupling limit, whereas our results are general and sustained beyond this limit. 

This paper is organised as follows. In Sec.~\ref{sec:cumulant_expansion} we formulate the grand potential in heavy quark QCD based on the cumulant expansion.
In Sec.~\ref{sec:cumulant} we calculate the quark-number susceptibilities up to the NLO in the HPE. We then investigate excitation energies of the quark and baryonic excitations in Sec.~\ref{sec:E}. We summarize our results in Sec.~\ref{sec:summary}.

\section{Grand potential in heavy-quark QCD}
\label{sec:cumulant_expansion}

\subsection{Lattice action}
\label{sec:action}

We consider SU(3) QCD on the lattice of size $N_x^3\times N_t$ with the $N_{\rm f}$-flavor Wilson fermion 
\begin{align}
    \label{eq:S_q}
    S_{\rm q}
    & =
    \sum^{N_{\rm f}}_{f=1}
    \sum_{s,s'}
    \bar{q}_{f}(s)
    \left[
        1-\kappa_{f} H(s, s')
    \right]
    q_{f}(s') , 
\end{align}
with the hopping matrix
\begin{align}
    \label{eq:hopping_matrix}
    H(s, s')
    \equiv & \,
    \sum_{k=1}^{3}
    \Big[
    \left(
        1 - \gamma_{k}
    \right)
    U_{k}(s) \delta_{s+\hat{k}, s'}
    \nonumber \\
    &
    \hspace{20pt}
    +
    \left(
        1 + \gamma_{k}
    \right)
    U^{\dagger}_{k}(s-\hat{k}) \delta_{s-\hat{k}, s'}
    \Big] \nonumber \\
    &
    \hspace{0pt}
    +
    \left[
    \left(
        1 - \gamma_{4}
    \right)
    e^{\mu_{f} a}
    U_{4}(s) \delta_{s+\hat{4}, s'}
    \right.
    \nonumber \\
    &
    \hspace{12pt}
    +
    \left.
    \left(
        1 + \gamma_{4}
    \right)
    e^{-\mu_{f} a}
    U^{\dagger}_{4}(s-\hat{4}) \delta_{s-\hat{4}, s'}
    \right],
\end{align}
where $U_\mu(s)$ denotes the link variable with $s$ being the lattice site, $a$ is the lattice spacing, and $\kappa_f$ and $\mu_f$ are the hopping parameter and the chemical potential for the $f$th flavor quark, respectively. 
The color and Dirac-spinor indices are suppressed for simplicity in Eq.~\eqref{eq:hopping_matrix}. 
The hopping parameters are related to the bare quark masses $m_f$ as $\kappa_{f} \simeq 1/(2 a m_{f})$ for large $m_f$~\cite{Rothe:1992nt}, and hence a small $\kappa_f$ corresponds to a heavy mass.
The chemical potentials are attached to the temporal links $U_{4}$ and $U_{4}^{\dagger}$ via exponential factors $e^{\pm\mu_{f} a}$~\cite{Hasenfratz:1983ce}. 
In the following, we assume the degenerated quark masses and the common chemical potential for all flavors, i.e., $\kappa_{f} = \kappa$ and $\mu_{f} = \mu_{\rm q}$, whereas the generalization to non-degenerate cases is straightforward.

The partition function of this system is expressed as 
\begin{align}
    Z
    & =
    \int \mathcal{D}U 
    \mathcal{D}\bar{q} \mathcal{D}q~
    e^{-S_{\rm g}-S_{\rm q}} 
    \notag \\
    & = 
    \int \mathcal{D} U
    e^{-S_{\rm g}
    + N_{\rm f} \ln {\rm Det} \left( 1 - \kappa H \right)
    }, 
    \label{eq:Z}
\end{align}
with the gauge action $S_{\rm g}$, where ${\rm Det}$ denotes the determinant over color, spinor, and spacetime indices.
The factor $N_{\rm f}$ in the exponent arises from a summation over $f$ in Eq.~\eqref{eq:S_q}. Throughout this study, the form of $S_{\rm g}$ is not specified, since the following discussion is independent of $S_{\rm g}$. On the Euclidean lattice, temperature is defined as $T \equiv 1/N_ta$.

Taking the heavy quark-mass limit, $\kappa \to 0$, QCD reduces to the ${\rm SU}(3)$ YM theory. This theory is known to exhibit a first-order phase transition at the critical temperature $T=T_{\rm c}$
characterized by the $\mathbb{Z}_3$ center symmetry of the ${\rm SU}(3)$ group.
The phase at $T > T_{\rm c}$ is referred to as the deconfined phase, where the center symmetry is spontaneously broken, while the symmetry is restored in the confined phase at $T < T_{\rm c}$. 
With finite $\kappa$, the fermion determinant breaks the $\mathbb{Z}_3$ symmetry explicitly, while the first-order phase transition continues to exist until a critical value is reached~\cite{Lo:2014vba,Cuteri:2020yke,Ashikawa:2024njc}. 

\subsection{Hopping parameter expansion}
\label{sec:HPE}

To investigate Eq.~\eqref{eq:Z} for small $\kappa$, it is convenient to adopt the Taylor expansion of the fermion determinant for $\kappa$ at $\kappa=0$~\cite{Rothe:1992nt},
\begin{align}
    N_{\rm f}\ln{\rm Det}\left( 1 - \kappa H \right)
    &=
    N_{\rm f}{\rm Tr}[ \ln( 1 - \kappa H) ] 
    \notag \\ 
    &= 
    - N_{\rm f} \sum^{\infty}_{l=1}
    \frac{\kappa^{l}}{l} {\rm Tr} [H^{l}],
    \label{eq:HPE}
\end{align}
where ${\rm Tr}$ denotes the trace over the same indices as ${\rm Det}$ in Eq.~\eqref{eq:Z}. Equation~\eqref{eq:HPE} is called the hopping-parameter expansion (HPE). 
Non-vanishing terms on the far right-hand side correspond to the closed trajectories on the lattice of length $l$ because of the trace operation for the spacetime indices.
Examples of the trajectories and their explicit calculation are found in Refs.~\cite{Kiyohara:2021smr,Ashikawa:2024njc,Rothe:1992nt,Wakabayashi:2021eye}.

The $\mu_{\rm q}$ dependence of each closed trajectory on the far  right-hand side of Eq.~\eqref{eq:HPE} are classified by the winding number $k$ in the temporal direction, where positive (negative) $k$ means the winding along the positive (negative) temporal direction. The terms with $k=0$, i.e., those without temporal windings, do not depend on $\mu_{\rm q}$, as they contain equal numbers of temporal links toward positive and negative directions, and the exponential factors $e^{\pm \mu_q a}$ cancel each other out.
In contrast, the terms with $k\ne0$ include $|k|N_t$ temporal links proportional to $e^{\pm \mu_q a}$ that do not cancel out. The terms with $k$ windings are thus proportional to $e^{k N_t \mu_{\rm q} a}
=e^{k\hat\mu_{\rm q}}$ with the dimensionless chemical potential $\hat\mu_{\rm q}\equiv \mu_{\rm q}/T$.

By factoring out the $\hat\mu_{\rm q}$ dependence, Eq.~\eqref{eq:HPE} is rewritten as
\begin{align}
    N_{\rm f}\ln {\rm Det}( 1 - \kappa H )
    &= \sum^{\infty}_{k=-\infty}
    e^{k \hat{\mu}_{\rm q}}  \Lambda_{k} ,
    \label{eq:HPEL} \\
    \Lambda_k
    &= - N_{\rm f}\sum_{\substack{ l\\\textrm{$k$ windings}}}
    \frac{\kappa^{l}}{l} {\rm Tr} [H^{l}]_{\mu_{\rm q}=0}.
    \label{eq:HPE2}
\end{align}
Notice that the $\kappa$ dependence is encoded in $\Lambda_k$.
From the definition, it immediately follows that $ \Lambda_{-k} =  \Lambda_{k}^{\dagger}$.
We notice that the fermionic action~\eqref{eq:S_q} depends on $e^{\hat\mu_{\rm q}}$ besides $\kappa$. Equation~\eqref{eq:HPEL} is the double expansion of Eq.~\eqref{eq:S_q} with respect to $e^{\hat\mu_{\rm q}}$ followed by $\kappa$.

The leading-order contributions of the HPE of $ \Lambda_k$ are calculated as follows. First, since $ \Lambda_0$ consists of loops without temporal windings, its LO term is given by 
\begin{align}
    \label{eq:Omega_0}
    \Lambda_{0}
    =
    -(2 \kappa)^{4} N_{\rm f}
    \sum_{s} \sum_{\mu < \nu} 
    {\rm Retr}_{\rm c}
    P_{\mu \nu}(s)
    +
    \mathcal{O}(\kappa^{6}),
\end{align}
with the plaquette
\begin{align}    
    \label{eq:plaquette}
    P_{\mu \nu}(s) 
    = U_{\mu}(s) U_{\nu}(s+\hat{\mu}) U^{\dagger}_{\mu}(s+\hat{\nu}) U^{\dagger}_{\nu}(s) ,
\end{align}
where ${\rm tr}_{\rm c}$ represents a trace over color indices and the NLO terms arise from the Wilson loops of length six, which are of order $\kappa^6$~\cite{Kiyohara:2021smr}. 
For $k \ne 0$, the shortest closed loop is the straight $k$-winding loop along the temporal direction, whose length is $|k|N_t$. The LO term in $\Lambda_k$ is thus of order $\kappa^{|k|N_t}$ and for $k > 0$ it reads~\cite{Wakabayashi:2021eye}
\begin{align}
    \label{eq:Lambda_k}
     \Lambda_k
    =
    (2 \kappa)^{k N_{t}}
    2 N_{\rm c}
    N_{\rm f}
    \frac{(-1)^{k+1}}{k}
    L_{k} 
    +
    \mathcal{O}(\kappa^{kN_{t}+2}) ,
\end{align}
where $N_{\rm c}=3$, and
\begin{align}
    \label{eq:Ploop_k}
    L_{k}
    &\equiv
    \frac1{N_{\rm c}}\sum_{\bm{s}}
    \mathrm{tr}_{\rm c}
    \left[
        \ell_{\rm P}(\bm{s})^{k}
    \right] ,
\end{align}
is the $k$-winding Polyakov loop with the local Polyakov-loop matrix
\begin{align}
    \label{eq:Ploop}
    \ell_{\rm P}(\bm{s})
    &\equiv
    {\cal P}     \prod^{N_{t}-1}_{j=0} U_{4}(\bm{s}+j\hat{4}) 
    \equiv U_4(\bm{s}) U_4(\bm{s}+\hat{4}) \cdots ,
\end{align}
at the space coordinate $\bm s$, where ${\cal P}$ means the path-ordered product as defined on the far right-hand side. 
The factor $(-1)^k$ in Eq.~\eqref{eq:Lambda_k} comes from the anti-periodic boundary conditions for the fermion fields.
The NLO terms in Eq.~\eqref{eq:Lambda_k} at order $\kappa^{kN_t+2}$ come from spatially bent Polyakov loops~\cite{Kiyohara:2021smr}.

Using Eq.~\eqref{eq:HPEL}, the partition function is represented as 
\begin{align}
    Z
    &= \int \mathcal{D} U e^{-S_{\rm g}}
    \exp\Bigg[
    \sum^{\infty}_{k=-\infty}
    e^{k \hat{\mu}_{\rm q}}  \Lambda_{k} \Bigg]
    \notag \\
    &= Z_{\rm g} \left\langle \exp\Bigg[
    \sum^{\infty}_{k=-\infty}
    e^{k \hat{\mu}_{\rm q}}  \Lambda_{k} \Bigg] \right\rangle ,
    \label{eq:ZL}
\end{align}
with $Z_{\rm g}= \int{\cal D}U e^{-S_{\rm g}}$ and the bracket denotes the expectation value defined by
\begin{align}
    \langle {\cal O} \rangle 
    = \lim_{\alpha\to0_{+}} \frac{1}{Z_{\rm g}} \int{\cal D}U e^{-S_{\rm g}-\alpha S_{\rm q}} {\cal O} ,
    \label{eq:<>}
\end{align}
where ${\cal O}$ is an operator composed of gauge degrees of freedom.
This expectation value reduces to that in the SU(3) YM theory, 
$\langle {\cal O} \rangle  = \int{\cal D}U e^{-S_{\rm g}} {\cal O}/Z_{\rm g}$, in the $\mathbb{Z}_3$ symmetric phase. 
In the $\mathbb{Z}_3$-broken phase, on the other hand, its meaning must be interpreted with care.
Since the fermion determinant in Eq.~\eqref{eq:Z} explicitly breaks the $\mathbb{Z}_3$ symmetry for finite $\kappa$, the expectation value must be taken in one of the three symmetry-broken vacua of YM theory, where $\langle\Lambda_k\rangle$ takes real values. As a result, the expectation values of $\mathbb{Z}_3$ variant operators in Eq.~\eqref{eq:ZL} can be nonzero in the broken phase.

\subsection{Cumulant expansion}
\label{sec:expansion}
The grand potential $\Omega(T,\mu_{\rm q}) $ is related to the partition function~\eqref{eq:ZL} as
\begin{align}
    \Omega(T,\mu_{\rm q}) 
    = - T \ln Z .
    \label{eq:Omega_def}
\end{align}
Applying the cumulant expansion introduced in App.~\ref{app:cumulant_exp}, Eq.~\eqref{eq:Omega_def} is calculated to be
\begin{align}
    \Omega (T,\mu_{\rm q}) 
    &= \Omega_{\rm g}(T)
    + \Omega_{\rm q}(T,\mu_{\rm q}) ,
    \label{eq:Omega_gq}
    \\
    \Omega_{\rm g} (T)
    &= -T \ln Z_{\rm g} ,
    \label{eq:Omega_g}
    \\
    -\frac{\Omega_{\rm q} (T,\mu_{\rm q}) }T
    &= \left\langle \prod^{\infty}_{k=-\infty} \left(
    \sum^{\infty}_{m_{k}=0}
    \frac{e^{k m_{k} \hat{\mu}_{\rm q}}}{m_{k}!} 
     \Lambda_{k}^{m_{k}} \right) \right\rangle_{\rm c}
    \notag \\
    &= \sum_{\{m_k\}} \frac{e^{\sum_{k} k m_k \hat\mu_{\rm q}}}{\prod_{k} m_k!} \left\langle \prod_{k} \Lambda_k^{m_k} \right\rangle_{\rm c} ,
    \label{eq:Omega_q}
\end{align}
where the cumulants $\langle \cdots \rangle_{\rm c}$ up to the third order are given by
\begin{align}
    \label{eq:cumulant1}
    \langle  \Lambda_i \rangle_{\rm c}
    & = 
    \langle  \Lambda_{i} \rangle,  
    \\
    \label{eq:cumulant2}
    \langle  \Lambda_i  \Lambda_j \rangle_{\rm c}
    & = 
    \langle  \Lambda_i  \Lambda_{j} \rangle
    -\langle  \Lambda_i \rangle \langle  \Lambda_j \rangle 
    \nonumber \\
    & = 
    \langle (  \Lambda_i - \langle  \Lambda_i \rangle )
    (  \Lambda_j - \langle  \Lambda_j \rangle ) \rangle ,
    \\
    \label{eq:cumulant3}
    \hspace{-8pt}
    \left \langle
         \Lambda_{i}
         \Lambda_{j}
         \Lambda_{k}
    \right \rangle_{\scriptsize{\rm c}}
    & = 
    \left \langle
        ( \Lambda_{i} -\langle \Lambda_{i} \rangle )
        ( \Lambda_{j} - \langle \Lambda_{j}  \rangle )
        ( \Lambda_{k} - \langle \Lambda_{k} \rangle ) 
    \right \rangle,
\end{align}
and the sum on the far right-hand side in Eq.~\eqref{eq:Omega_q} is taken over all possible combinations of non-negative integers $m_k$. 
See App.~\ref{app:cumulant_exp} 
for the derivation of Eq.~\eqref{eq:Omega_q} and the general definition of the cumulants.

Equation~\eqref{eq:Omega_q} can be rewritten as
\begin{align}
    \frac{- \Omega_{\rm q}(T, \mu_{\rm q})}{T}
    &= 
    X_{0} + \sum^{\infty}_{w=1} 
    \left( e^{w \hat{\mu}_{\rm q}} X_w + e^{-w \hat{\mu}_{\rm q}} X_{-w} \right) 
    \notag \\
    &= 
    X_{0} + \sum^{\infty}_{w=1}     \left( e^{w \hat{\mu}_{\rm q}} + e^{-w \hat{\mu}_{\rm q}} \right) X_w ,
    \label{eq:grand_pot_X}
\end{align}
where $X_w$ contains the terms satisfying $w = \sum_k k m_k$. On the second equality of Eq.~\eqref{eq:grand_pot_X}, we used $X_{-w}=X^*_w=X_w$, which follows from the time reversal symmetry. Examples of the terms included in $X_0$ and $X_1$ are,
\begin{align}
    X_0 
    &= \langle \Lambda_0 \rangle_{\rm c}
    + \frac12 \langle \Lambda_0^2 \rangle_{\rm c}
    + \langle \Lambda_1 \Lambda_{-1} \rangle_{\rm c} 
    + \frac12 \langle \Lambda_1^2 \Lambda_{-2} \rangle_{\rm c} 
    + \cdots,
    \\
    X_1 
    &= \langle \Lambda_1 \rangle_{\rm c}
    + \langle \Lambda_0 \Lambda_1 \rangle_{\rm c}
    + \langle \Lambda_2 \Lambda_{-1} \rangle_{\rm c}
    + \cdots .
\end{align}
Since the leading-order terms of $X_0$ and $X_1$ come from $\langle \Lambda_0 \rangle_{\rm c}$ and $\langle \Lambda_1 \rangle_{\rm c}$, respectively, their HPEs are given by
\begin{align}
    \label{eq:X0}
    X_{0} 
    &= N_{\rm f} (2 \kappa)^{4} 
    \sum_{s} \sum_{\mu < \nu} 
    \left \langle \mathrm{Re tr}_{\rm c} P_{\mu \nu}(s) \right \rangle
    + \mathcal{O}(\kappa^{6}) , 
    \\
    \label{eq:X1}
    X_{1}
    & = 
    2 N_{\rm c} N_{\rm f} (2 \kappa)^{N_{t}} \langle L_1 \rangle
    + \mathcal{O}(\kappa^{N_{t}+2}) .
\end{align}
A similar manipulation for $w=2,3$ leads to 
\begin{align}
    \label{eq:X2}
    X_{2}
    & = \langle \Lambda_2 \rangle_{\rm c}
    + \frac12 \langle \Lambda_1^2 \rangle_{\rm c}
    + {\cal O}(\kappa^{2N_t+4})
    \notag \\
    &=
    (2 \kappa)^{2 N_{t}}
    \Big\{ -N_{\rm c} N_{\rm f} \langle L_{2} \rangle + 2 N_{\rm c}^2 N_{\rm f}^2 \langle L_{1}^2 \rangle_{\rm c} \Big\} 
    \notag \\
    &\mathrel{\phantom=}
    + \mathcal{O}(\kappa^{2 N_{t}+2}) , 
    \\
    \label{eq:X3}
    X_3
    &= \langle \Lambda_3 \rangle_{\scriptsize{\rm c}}
    + \langle \Lambda_2\Lambda_1 \rangle_{\rm c}
    + \frac1{3!} \langle \Lambda_1^3 \rangle_{\rm c}
    + \mathcal{O}(\kappa^{3 N_{t}+4}) 
    \nonumber \\
    & = 
    (2 \kappa)^{3 N_t}
    \bigg\{ 
        \frac{2 N_{\rm c} N_{\rm f}}3 \langle L_3 \rangle 
        - 2N_{\rm c}^2 N_{\rm f}^2 \langle L_2 L_1 \rangle_{\rm c}
        \notag \\
        & \hspace{40pt}
        + \frac{4 N_{\rm c}^3 N_{\rm f}^3}3 \langle L_1^3 \rangle_{\rm c}
    \bigg\} 
    +\mathcal{O}(\kappa^{3 N_{t}+2}).
\end{align}
As found in Eqs.~\eqref{eq:X1}--\eqref{eq:X3}, $X_{w}~(w \ge 1)$ is of order $\kappa^{w N_{t}}$,
while $X_0$ is of order $\kappa^4$.

We notice that $X_w$ consists of the expectation values of $\mathbb{Z}_3$ invariant operators when $w$ is a multiple of three, $w\in3\mathbb Z$, while in other cases, all operators are not $\mathbb{Z}_3$ invariant. 
In the confined phase where the $\mathbb{Z}_3$ symmetry remains unbroken, expectation values of all operators that are not $\mathbb{Z}_3$ invariant must vanish, and we obtain $X_w=0$ unless $w$ is a multiple of three. In contrast, $X_w$ can be nonzero for all $w$ in the deconfined phase where the $\mathbb{Z}_3$ symmetry is spontaneously broken.
These properties of $X_w$ play a crucial role in later sections.

As the final remark of this section, we notice that $X_w$ are extensive variables, i.e., they are proportional to the spatial volume $V=N_{x}^{3}a^{3}$. This property is guaranteed by the extensivity of cumulants~\cite{Asakawa:2015ybt}. In fact, 
$\langle L_2 L_1 \rangle_{\rm c}$ in Eq.~\eqref{eq:X3}, for example, is rewritten using the translational symmetry as
\begin{align}
    \langle L_2 L_1 \rangle_{\rm c}
    &= \frac1{N_{\rm c}^2} \sum_{\bm{s},\bm{s}'} \langle {\rm tr}[\ell_{\rm P}(\bm{s})^2] {\rm tr}\ell_{\rm P}(\bm{s}') \rangle_{\rm c}
    \notag \\
    &= \frac{N_x^3}{N_{\rm c}^2} \sum_{\bm{s}} \langle {\rm tr}[\ell_{\rm P}(\bm{s})^2] {\rm tr}\ell_{\rm P}(\bm{0}) \rangle_{\rm c} .
    \label{eq:Lambda21}
\end{align}
Because $\langle {\rm tr}[\ell_{\rm P}(\bm{s})^2] {\rm tr}\ell_{\rm P}(\bm{0}) \rangle_{\rm c}$ will damp quickly at large $|\bm s|$, the sum over $\bm s$ in Eq.~\eqref{eq:Lambda21} gives a finite value that does not depend on $V$. Equation~\eqref{eq:Lambda21} thus is proportional to $V$.

\section{Quark-number susceptibility}
\label{sec:cumulant}

Now, let us investigate the thermal properties of heavy-quark QCD using the grand potential~\eqref{eq:grand_pot_X}.
In this section, we focus on the quark-number susceptibilities
\begin{align}
    \chi_n (T,\mu_{\rm q})
    \equiv
    - \frac{\partial^{n}}{\partial \hat{\mu}_{\rm q}^{n}} 
    \frac{\Omega(T, \mu_{\rm q})}{TV}
    \bigg|_T 
    =
    - \frac{\partial^{n}}{\partial \hat{\mu}_{\rm q}^{n}} 
    \frac{\Omega_{\rm q}(T, \mu_{\rm q})}{TV}
    \bigg|_T ,
    \label{eq:chi}
\end{align}
which are related to the quark-number cumulants as 
\begin{align}
    \label{eq:Nqc}
    \langle N_{\rm q}^n \rangle_{\rm c} 
    &= V\chi_n (T,\mu_{\rm q}),
\end{align}
where the $\langle N_{\rm q}^n \rangle_{\rm c}$ are defined similarly to Eqs.~\eqref{eq:cumulant1}--\eqref{eq:cumulant3} with a replacement of $\Lambda$'s with the quark number $N_q$~\cite{Asakawa:2015ybt}.

It is known that $\chi_n (T,\mu_{\rm q})$ are sensitive to the quark charge carried by quasi-particle excitations in the system~\cite{Ejiri:2005wq,Asakawa:2015ybt}, especially when they are well described by the Boltzmann statistics. In this case, and when all the quasi-particles in the system carry an identical quark charge $Q$, the ratios of $\chi_n (T,\mu_{\rm q})$ satisfy simple relations 
\begin{align}
    \frac{\chi_{n+n'} (T,\mu_{\rm q})}{\chi_n (T,\mu_{\rm q})}
    =\frac{\langle N_{\rm q}^{n+n'} \rangle_{\rm c}}{\langle N_{\rm q}^n \rangle_{\rm c}}=Q^{n'} ,
\end{align}
for positive and even $n'\in2\mathbb{N}$ unless the denominators vanish. 

\subsection{Leading order}

The quark-number susceptibilities in heavy-quark QCD are calculated from Eqs.~\eqref{eq:Omega_q} and~\eqref{eq:chi} as
\begin{align}
    \label{eq:cumulant_n_hpe}
    \chi_n(T,\mu_{\rm q})
    &= \frac1V
    \sum^{\infty}_{w=1} w^{n}
    C_{w,n} X_{w} , 
    \\
    \label{eq:Cw}
    C_{w,n}
    &= e^{w\hat\mu_{\rm q}}+(-1)^{n} e^{-w\hat\mu_{\rm q}},
\end{align}
where $C_{w,n+n'}=C_{w,n}$ for $n'\in2\mathbb{N}$. 

In Eq.~\eqref{eq:cumulant_n_hpe}, the order of $X_w$ in the HPE is $\kappa^{wN_t}$. Therefore, the more dominant contributions in the HPE are contained in $X_w$ at smaller $w$. However, the terms with larger $w$ can give a larger contribution for large $\mu_{\rm q}$ when $e^{|\hat\mu_{\rm q}|}\kappa^{N_t}>1$ even for small $\kappa$. This is related to the fact that $e^{\pm\hat\mu_{\rm q}}\kappa$ are expansion parameters of the action~\eqref{eq:HPE} besides $\kappa$. In the following, we assume $e^{|\hat\mu_{\rm q}|}\kappa^{N_t}\ll 1$ besides $\kappa\ll 1$ to ensure the convergence of Eq.~\eqref{eq:cumulant_n_hpe}.

At $T>T_{\rm c}$, because of the spontaneous breaking of the $\mathbb{Z}_3$ symmetry, the leading-order terms in Eq.~\eqref{eq:cumulant_n_hpe} come from $w=1$ at order $\kappa^{N_t}$, 
\begin{align}
    \chi_n(T,\mu_{\rm q})
    = \frac1V 
    C_{1,n}X_1 
    + \mathcal{O}(\kappa^{2N_{t}}) ,
    \label{eq:chi_n_leading}
\end{align}
where $\mathcal{O}(\kappa^{2N_{t}})$ come from the terms with $w\ge2$. Equation~\eqref{eq:chi_n_leading} shows that 
\begin{align}
    \label{eq:cumulant_ratio_quark}
    \chi_{n+n'}(T,\mu_{\rm q}) = \chi_n(T,\mu_{\rm q})
    \qquad\textrm{up to order $\kappa^{2N_t-1}$},
\end{align}
for $n'\in 2\mathbb{N}$.
For $T<T_{\rm c}$, on the other hand, the leading-order terms arise from $w=3$ owing to the $\mathbb{Z}_3$ symmetry, which leads to 
\begin{align}
    \label{eq:cumulant_ratio_hadron}
    \chi_{n+n'}(T,\mu_{\rm q}) = 3^{n'} \chi_n(T,\mu_{\rm q})
    \qquad\textrm{up to order $\kappa^{6N_t-1}$},
\end{align}
From Eqs.~\eqref{eq:cumulant_ratio_quark} and~\eqref{eq:cumulant_ratio_hadron}, we obtain 
\begin{eqnarray}
    \label{eq:QNS_summary}
    \frac{\chi_{n+n'} (T,\mu_{\rm q})}{\chi_n (T,\mu_{\rm q})}
    = \frac{\langle N_{\rm q}^{n+n'} \rangle_{\rm c} }{ \langle N_{\rm q}^n \rangle_{\rm c} }
    =
    \begin{cases}
    1      & (T>T_{\rm c}), \\
    3^{n'} & (T<T_{\rm c}),
\end{cases}
\end{eqnarray}
up to order $\kappa^{2N_t-1}$ in the HPE, except for the case where both the denominator and numerator vanish.
This relation \eqref{eq:QNS_summary} also holds in the finite density system.

The result in Eq.~\eqref{eq:QNS_summary} indicates that the quark charges carried by elementary excitations are $\pm1$ and $\pm3$ above and below $T_{\rm c}$, respectively~\cite{Ejiri:2005wq}. In other words, the quarks are confined into baryonic excitations below $T_{\rm c}$, while excitations above $T_{\rm c}$ are dominated by single-quark excitations. It is notable that Eq.~\eqref{eq:QNS_summary} is obtained as a direct consequence of the $\mathbb{Z}_3$ symmetry. We will pursue this picture for further exploration in the next section.

\subsection{next-to-leading order}
\label{sec:NLO}
We can extend the analysis of the ratios  $\chi_{n+n'}/\chi_n$ to NLO. In the deconfined phase for $T>T_{\rm c}$, from Eq.~\eqref{eq:cumulant_n_hpe} the ratio up to the NLO is calculated to be
\begin{align}
    \frac{\chi_{n+n'}(T,\mu_{\rm q})}{\chi_n(T,\mu_{\rm q})}
    &=
    1 + (2^{n+n'}-2^n)
    \frac{C_{2,n+n'}}{C_{1,n}}
    \frac{X_2}{X_1}  + \mathcal{O}(\kappa^{2N_t}) ,
    \label{eq:NLO1}
    \\
    \frac{X_2}{X_1}
    &=
    (2\kappa)^{N_t}
    \frac{N_{\rm c}N_{\rm f}\langle L_1^2 \rangle_{\rm c}-\langle L_2 \rangle}{2\langle L_1 \rangle}   + \mathcal{O}(\kappa^{N_t+2}),
    \label{eq:X2/X1}
\end{align}
for $n' \in 2\mathbb{N}$, where $X_2/X_1$ is of order $\kappa^{N_t}$. It is worth emphasizing that the NLO of $\chi_n(T,\mu_{\rm q})$ is only of order $\kappa^2$ higher than the LO terms. However, all terms in $X_1$ cancel with one another between the numerator and denominator in the ratio, and this enhances the order of the NLO terms significantly.

In the confined phase below $T_{\rm c}$, from Eq.~\eqref{eq:cumulant_ratio_hadron} the NLO contributions are given by the ratio between $X_6$ and $X_3$ as
\begin{widetext}
\begin{align}
    &\frac{\chi_{n+n'}(T,\mu_{\rm q})}{\chi_n(T,\mu_{\rm q})}
    =
    3^{n'} + 3^{n'}(2^{n+n'}-2^n)
    \frac{C_{6,n+n'}}{C_{3,n}}
    \frac{X_6}{X_3}+\mathcal{O}(\kappa^{6N_t}),
    \\
    \label{eq:nume_X6X3}
    &X_6
    =(2\kappa)^{6N_t}\Bigg[
    -\frac{N_{\rm c}N_{\rm f}}{3}\langle L_6\rangle
    +N_{\rm c}^2N_{\rm f}^2\bigg( \frac{4}{5}\langle L_5L_1\rangle_{\rm c} +\frac{1}{2}\langle L_4L_2\rangle_{\rm c} +\frac{2}{9}\langle L_3^2\rangle_{\rm c}\bigg)
    -N_{\rm c}^3N_{\rm f}^3\bigg(\langle L_4L_1^2\rangle_{\rm c} +\frac43\langle L_3L_2L_1\rangle_{\rm c} +\frac{1}{6}\langle L_2^3\rangle_{\rm c}\bigg) \notag \\
    &\hspace{55pt}
    +N_{\rm c}^4 N_{\rm f}^4 \bigg( \frac{8}{9}\langle L_3 L_1^3 \rangle_{\rm c} +\langle L_2^2 L_1^2\rangle_{\rm c}\bigg)
    -\frac{2}{3} N_{\rm c}^5 N_{\rm f}^5 \langle L_2L_1^4\rangle_{\rm c}
    +\frac{4}{45} N_{\rm c}^6 N_{\rm f}^6 \langle L_2L_1^4\rangle_{\rm c}\Bigg]+\mathcal{O}(\kappa^{6N_t+2}),
\end{align}
\end{widetext}
where the order of the NLO terms is $\kappa^{3N_t}$.

\section{Excitation energies}
\label{sec:E}

Next, we take a closer look at the properties of Eq.~\eqref{eq:Omega_q} in connection with elementary excitations in the deconfined and confined phases.

\subsection{Lattice Boltzmann gas}
\label{sec:latticegas}

For this purpose, we first consider the thermodynamics of a lattice gas of heavy-mass fermions with degeneracy $g$. When their mass is sufficiently large, fermionic excitations on the lattice occur on a single lattice site without hopping to neighboring sites. The quantum state of a single lattice site is specified solely by the existence of excited fermions, and its partition function is given by
\begin{align}
    Z_{\rm single} 
    = \left(1 + e^{-(E-q\mu)/T} \right)^g,
    \label{eq:Zsingle}
\end{align}
where $E$ and $q$ are the excitation energy and the charge of the fermion, and $\mu$ is the chemical potential of the charge.

When the hopping of fermions is negligible, one can also assume that the quantum states of individual lattice sites are independent of one another. 
The partition function of the total system of volume $N_x^3$ thus reads $Z_{\rm total} = Z_{\rm single}^{N_x^3}$, and the grand potential is calculated to be
\begin{align}
    \Omega_{\rm total} 
    &= -T \ln Z_{\rm total} 
    = -T N_x^3 \ln Z_{\rm single} 
    \notag \\
    &= -gT N_x^3 \ln \left(1+e^{-(E-q\mu)/T}\right)
    \notag \\
    &\simeq -g T N_x^3 e^{-(E-q\mu)/T},
    \label{eq:Omega_heavy}
\end{align}
where in the last line we used an approximation $\ln(1+x)\simeq x$ for small $x$, which is well justified for $E-q\mu\gg T$ corresponding to the Boltzmann statistics.

In relativistic systems, all particles have their associated antiparticles with opposite charges. The grand potential in this case thus reads
\begin{align}
    \Omega_{\rm total} 
    = -g T N_x^3 ( e^{q\mu/T} + e^{-q\mu/T} ) e^{-E/T}.
    \label{eq:Omega_heavyR}
\end{align}

Notice that the derivatives of Eq.~\eqref{eq:Omega_heavy} give $\chi_{n+n'} /\chi_n=q^{n'}$. This result is consistent with Eq.~\eqref{eq:QNS_summary} with $q=1$ and $3$ for deconfined and confined phases, respectively. This agreement shows that the fermionic excitations in heavy-quark QCD are indeed the lattice gas in Eq.~\eqref{eq:Omega_heavyR}. In what follows, using this correspondence, we evaluate the excitation energies above and below $T_{\rm c}$.

\subsection{Quark excitation energy at $T>T_{\rm c}$}
\label{sec:E>}

The grand potential~\eqref{eq:Omega_gq} in the HPE at $T>T_{\rm c}$ reads
\begin{align}
    \Omega 
    &= \Omega^{(0)} + \Omega^{(1)} + {\cal O}(\kappa^{2N_t}),
    \label{eq:Omega01}
    \\
    \Omega^{(0)} 
    &= -T\left( \ln Z_{\rm g} + X_0 \right),
    \label{eq:Omega0}
    \\
    \Omega^{(1)}
    &= -T ( e^{ \hat{\mu}_{\rm q}} + e^{-\hat{\mu}_{\rm q}} ) X_1
    \notag \\
    &= 
    -2N_{\rm c} N_{\rm f} T N_x^3 
    (e^{\hat\mu_{\rm q}} + e^{-\hat\mu_{\rm q}})  (2\kappa)^{N_t}\frac{\langle L_1\rangle}{N_x^3}
    \notag \\
    & \mathrel{\phantom=}
    + {\cal O}(\kappa^{N_t+2}),
    \label{eq:Omega1} 
\end{align}
where $\Omega^{(0)}$ represents the $\mu_{\rm q}$-independent term, and $\Omega^{(1)}$ is the leading-order term having $\mu_{\rm q}$ dependence. 
Comparing Eq.~\eqref{eq:Omega1} with Eq.~\eqref{eq:Omega_heavyR}, one finds that $\Omega^{(1)}$ is regarded as the thermodynamic potential of the lattice Boltzmann gas with $g=2N_{\rm c}N_{\rm f}$, $q=1$, and the excitation energy
\begin{align}
    E &= -T\ln \bigg[ (2\kappa)^{N_t} \frac{\langle L_1\rangle}{N_x^3} \bigg]
    \notag \\
    &=
    -T \ln \bigg[ (2\kappa)^{N_t} \frac1{N_{\rm c}N_x^3} \sum_{\bm s} \langle {\rm tr}_{\rm c} \ell_{\rm P} (\bm s)\rangle\bigg] .
    \label{eq:E_quark}
\end{align} 
From the fact that the pole mass $m_{\rm pole}$ of the free Wilson fermion is related to the hopping parameter as~\cite{Rothe:1992nt}
\begin{equation}
    \label{eq:pole_mass}
    m_{\rm pole} a
    = \ln \left[ 1 + \frac12 \Big(
        \frac{1}{\kappa} - \frac{1}{\kappa_{\rm c}}
        \Big) \right] ,
\end{equation}
with $\kappa_{\rm c}=1/8$ for free fermions, we obtain $\ln[(2\kappa)^{N_t}] = -m_{\rm pole}/T$ in the heavy-mass limit $\kappa\to0$. Equation~\eqref{eq:E_quark} is thus rewritten as
\begin{align}
    E = m_{\rm pole} - T \ln 
    \bigg[
    \frac1{N_{\rm c}N_x^3} \sum_{\bm s} \langle {\rm tr}_{\rm c} \ell_{\rm P} (\bm s)\rangle \bigg] . 
    \label{eq:E_quark2}
\end{align}
Since the second term is the free energy of a static quark on the lattice~\cite{Rothe:1992nt}, Eq.~\eqref{eq:E_quark2} is a reasonable result to state that the excitation energy of a quark is given by the sum of the pole mass and the free energy.
The degeneracy factor $g=2N_{\rm c}N_{\rm f}$, of course, comes from the spin, color, and flavor degrees of freedom of quarks. We notice that $\langle {\rm tr}_{\rm c}\ell_{\rm P}\rangle/N_{\rm c}\le1$ and thus the second term in Eq.~\eqref{eq:E_quark2} is positive.

In Eq.~\eqref{eq:Omega0}, $X_0$ is interpreted as the modification of the grand potential of the gluonic medium due to an interplay with heavy quarks and the contribution of the mesonic excitations. We notice that the order of $X_0$ is $\kappa^4$, which is parametrically smaller than that of $X_1$. This means that the latter is invisible in the simple HPE of $\Omega$. Nevertheless, we can obtain excitation properties of quarks, Eq.~\eqref{eq:E_quark2}, by focusing on the $\mu_{\rm q}$ dependence.

\subsection{Baryonic excitation energy at $T<T_{\rm c}$}
\label{sec:E<}

Next, we apply the same argument to the grand potential at $T<T_{\rm c}$. In this case, owing to the $\mathbb{Z}_3$ symmetry we obtain
\begin{align}
    \Omega 
    &= \Omega^{(0)} + \Omega^{(3)} + {\cal O}(\kappa^{6N_t})
    \label{eq:Omega03} \\
    \Omega^{(3)}
    &= -T ( 
    e^{ 3\hat{\mu}_{\rm q}} + e^{-3\hat{\mu}_{\rm q}} ) X_3
    \notag \\
    &= - T N_x^3
    (2\kappa)^{3N_t}  (e^{3\hat\mu_{\rm q}} + e^{-3\hat\mu_{\rm q}}) J  + {\cal O}(\kappa^{3N_t+2}) ,
    \label{eq:Omega3} \\
    J
    &= \frac{2N_{\rm c}N_{\rm f}}{3}\frac{
        \langle L_3 \rangle
        }{N_x^3}
        -
        2 N_{\rm c}^2 N_{\rm f}^{2}
        \frac{
        \langle L_{2} L_{1} \rangle_{\rm c}
        }{N_x^3}
        +
        \frac{4 N_{\rm c}^3 N_{\rm f}^{3}}{3}
        \frac{
        \langle L_{1}^{3} \rangle_{\rm c}}{N_x^3}.
    \label{eq:J}
\end{align}
Here, $\mu_{\rm q}$ dependence of $\Omega^{(3)}$ shows that this term is the contribution of excitations having three quark charges, i.e., baryonic states, to the thermodynamic potential.

One subtlety in applying the same argument as the previous subsection to Eq.~\eqref{eq:Omega3} is that the baryonic excitations have many resonance states, whose energy spectrum would not be well separated even for heavy-quark masses; see, for example, Ref.~\cite{Yu:2025gdg} and references therein. Here, we denote their excitation energies and degeneracy factors as $E_b$ and $g_b$, respectively, where the subscript $b$ runs over the ground and all resonance states. Following the hadron resonance gas model~\cite{Braun-Munzinger:2003pwq}, we then assume that the thermodynamic potential is given by the sum of those of resonances, which are treated as independent free lattice gas. Since $E_b\gtrsim 3m_{\rm pole}$ is large for $\kappa\to0$, the contribution of the baryonic excitations to the grand potential in this picture reads
\begin{align}
    \Omega_{\rm B} = - T N_x^3 ( e^{3\hat\mu_{\rm q}} + e^{-3\hat\mu_{\rm q}}) \sum_{b\in\textrm{(baryons)}} g_b e^{-E_b/T} .
    \label{eq:Omega_B}
\end{align}
Comparing Eq.~\eqref{eq:Omega_B} with Eqs.~\eqref{eq:Omega_heavyR} and~\eqref{eq:Omega3}, one finds
\begin{align}
    \sum_b g_b e^{-(E_b-3m_{\rm pole})/T} 
    = J,
    \label{eq:ExB}
\end{align}
where we used $m_{\rm pole}/T=-N_t \ln(2\kappa)$.
In Eq.~\eqref{eq:ExB}, $E_b-3m_{\rm pole}$ is interpreted as the binding energy of the baryon $b$. 

To gain further insight into the baryon spectrum, we now focus on the $N_{\rm f}$ dependence of Eq.~\eqref{eq:J}, which is given by the polynomial of $N_{\rm f}$ up to the third order. To understand this $N_{\rm f}$ dependence, we remark that all baryonic states in our system belong to one of the flavor multiplets of the flavor ${\rm SU}(N_{\rm f})$ symmetry. The direct product of flavor composition of baryons is decomposed as 
\begin{align}
    \bm{N}_{\rm f} \otimes \bm{N}_{\rm f} \otimes \bm{N}_{\rm f}
    = \bm{R}_{\rm S} \oplus \bm{R}_{\rm MS} \oplus \bm{R}_{\rm MA} \oplus \bm{R}_{\rm A},
    \label{eq:flavor_sum}
\end{align}
where $\bm{R}_{\rm S}$, $\bm{R}_{\rm MS}$, $\bm{R}_{\rm MA}$, and $\bm{R}_{\rm A}$ represent the completely-symmetric, mixed-symmetric, mixed-anti-symmetric, and completely-anti-symmetric representations, respectively. 
Their dimensions are~\cite{Georgi:2000vve}
\begin{align}
    d_{\rm S} &= {}_{N_{\rm f}+2}C_3,
    &
    d_{\rm A} &= {}_{N_{\rm f}}C_3 ,
    \label{eq:d_r}
\end{align}
for $\bm{R}_{\rm S}$ and $\bm{R}_{\rm A}$, and
\begin{align}
    d_{\rm MS} = d_{\rm MA} \equiv d_{\rm M} = 2_{N_{\rm f}+1}C_3 ,
    \label{eq:d_r}
\end{align}
for $\bm{R}_{\rm MS}$ and $\bm{R}_{\rm MA}$. 
One can thus decompose the sum over all baryons in Eq.~\eqref{eq:ExB} into those belonging to each flavor multiplet as
\begin{align}
    J_r = \sum_{b\in r} g_b e^{-(E_b-3m_{\rm pole})/T} && (r={\rm S,M,A}) ,
    \label{eq:ExBr}
\end{align}
where we do not distinguish MS and MA because they form a single multiplet together with the mixed symmetry of other degrees of freedom~\cite{Georgi:2000vve}.

To understand the $N_{\rm f}$ dependence of Eq.~\eqref{eq:ExBr}, we point out that the excitation energies $E_b$, as well as the spin and angular-momentum structure, in each multiplet should be independent of $N_{\rm f}$. This statement would be justified by the fact that baryonic states in our system are excited so rarely that they can be regarded as non-interacting with one another. Hence, their properties should be determined solely by the flavor composition in a single excitation. This means that $e^{-(E_b-3m_{\rm pole})/T}$ is $N_{\rm f}$ independent. On the other hand, the degeneracy $g_b$ in the representation $\bm{R}_r$ must be proportional to $d_r$ owing to the flavor symmetry. Therefore, $N_{\rm f}$ dependence of Eq.~\eqref{eq:ExBr} comes only from $d_r$, while $J_r/d_r$ is $N_{\rm f}$ independent. Rewriting Eq.~\eqref{eq:ExB} as 
\begin{align}
    J &= d_{\rm S} \frac{J_{\rm S}}{d_{\rm S}} + d_{\rm M} \frac{J_{\rm M}}{d_{\rm M}} + d_{\rm A} \frac{J_{\rm A}}{d_{\rm A}} 
    \notag \\
    &= {}_{N_{\rm f}+2}C_3 \frac{J_{\rm S}}{d_{\rm S}} + 2_{N_{\rm f}+1}C_3 \frac{J_{\rm M}}{d_{\rm M}} + {}_{N_{\rm f}}C_3 \frac{J_{\rm A}}{d_{\rm A}} ,
    \label{eq:Jdecomp}
\end{align}
and comparing the $N_{\rm f}$ dependence of Eq.~\eqref{eq:Jdecomp} with that of Eq.~\eqref{eq:J}, one finds that $J_r/d_r$ are given by
\begin{align}
    \frac{J_{\rm S}}{d_{\rm S}}
    &= \frac{2N_{\rm c}}3 \frac{\langle L_3 \rangle}{N_x^3} - 2N_{\rm c}^2 \frac{\langle L_2 L_1 \rangle_{\rm c}}{N_x^3} + \frac{4N_{\rm c}^3}3 \frac{\langle L_1^3 \rangle_{\rm c}}{N_x^3} ,
    \label{eq:J_S} \\
    \frac{J_{\rm M}}{d_{\rm M}}
    &= -\frac{2N_{\rm c}}3 \frac{\langle L_3 \rangle}{N_x^3} + \frac{8N_{\rm c}^3}3 \frac{\langle L_1^3 \rangle_{\rm c}}{N_x^3} ,
    \label{eq:J_MS} \\
    \frac{J_{\rm A}}{d_{\rm A}}
    &= \frac{2N_{\rm c}}3 \frac{\langle L_3 \rangle}{N_x^3} + 2N_{\rm c}^2 \frac{\langle L_2 L_1 \rangle_{\rm c}}{N_x^3} + \frac{4N_{\rm c}^3}3 \frac{\langle L_1^3 \rangle_{\rm c}}{N_x^3} .
    \label{eq:J_A}
\end{align}

In Eqs.~\eqref{eq:J_S}--\eqref{eq:J_A}, we have succeeded in decomposing the grand potential into those of individual flavor multiplets.
It is worthwhile to notice that the contribution from the completely anti-symmetric channel is mandatory in this decomposition. This result is interesting because there are no candidates for such baryons in the experimentally observed baryons in the light-quark sector composed of $u,d,s$ quarks~\cite{ParticleDataGroup:2024cfk}. 

Finally, to make a qualitative estimate of the values of Eqs.~\eqref{eq:J_S}--\eqref{eq:J_A}, let us calculate them in the strong-coupling limit of the gluonic action $S_{\rm g}$. In this case, all link variables are independent and their distribution is uniform in the Haar space of SU(3). Therefore, $\langle L_2 L_1 \rangle_{\rm c}$, and $\langle L_1^3 \rangle_{\rm c}$ are given by the spatially overlapping terms, i.e.,
\begin{align}
    \frac{\langle L_2 L_1 \rangle_{\rm c} }{N_x^3} 
    &= \frac1{N_{\rm c}^2 N_{x}^3} \sum_{\bm{s}} \langle {\rm tr}_{\rm c} [\ell_{\rm P}(\bm s)^2] {\rm tr}_{\rm c} \ell_{\rm P}(\bm s)\rangle ,
    \\
    \frac{\langle L_1^3 \rangle_{\rm c} }{N_x^3} 
    &= \frac1{N_{\rm c}^3 N_{x}^3} \sum_{\bm{s}} \langle ({\rm tr}_{\rm c} \ell_{\rm P}(\bm s))^3 \rangle .
\end{align}
Moreover, because of the uniform distribution in the Haar space, their expectation values are analytically calculated using the identity
\begin{align}
    \int dU U_{aa'} U_{bb'} U_{cc'} = \frac1{3!}\varepsilon_{abc}\varepsilon_{a'b'c'} ,
\end{align}
for $U \in {\rm SU(3)}$ as 
\begin{align}
    \frac{\langle L_3 \rangle}{N_x^3} 
    &= \frac1{N_{\rm c}} ,
    &
    \frac{\langle L_2 L_1 \rangle_{\rm c} }{N_x^3} 
    &= -\frac1{N_{\rm c}^2} ,
    &
    \frac{\langle L_1^3 \rangle_{\rm c} }{N_x^3} 
    &= \frac1{N_{\rm c}^3} .
\end{align}
Substituting these values into Eqs.~\eqref{eq:J_S}--\eqref{eq:J_A} yields
\begin{align}
    \frac{J_{\rm S}}{d_{\rm S}} = 4, &&
    \frac{J_{\rm M}}{d_{\rm M}} = 2, &&
    \frac{J_{\rm A}}{d_{\rm A}} = 0.
    \label{eq:Jest}
\end{align}
Therefore, $J_{\rm A}$ vanishes in this case. The values of $J_{\rm S}$ and $J_{\rm M}$ suggest that there exists only a single baryonic state with $E_{\rm b}-m_{\rm pole}=0$ in these channels with the degeneracy factors $g_{\rm S}=4d_{\rm S}$ and $g_{\rm M}=2d_{\rm M}$, respectively. The factors four and two are interpreted as the spin degeneracy, 
since the spins of color-singlet zero angular-momentum baryons
in S and M multiplets are $s=3/2$ and $1/2$, respectively, from the Pauli principle.

We emphasize that Eq.~\eqref{eq:Jest} is the result obtained in the strong coupling limit. Our results Eqs.~\eqref{eq:J_S}--\eqref{eq:J_A} are not constrained by this condition when the expectation values are calculated in SU(3) YM theory. It is worthwhile mentioning that the local Polyakov-loop matrix $\ell_{\rm P}(\bm s)$ in the confined phase is known to be almost randomly distributed in the Haar space~\cite{Hanada:2023krw,Hanada:2023rlk}. This suggests that Eq.~\eqref{eq:Jest} may give a good estimate even in more realistic situations. To obtain quantitative results, we need the Monte Carlo simulations of the expectation values, which are left for future study.




\section{Summary}
\label{sec:summary}

In this paper, we have investigated the thermodynamic properties of excitations carrying quark charges in heavy-quark QCD based on the hopping parameter expansion (HPE). We have calculated the grand potential in the HPE with the aid of the cumulant expansion. We derived an analytic expression of the grand potential classified by the chemical potential dependence. 

Using the grand potential thus obtained, we have computed the quark-number susceptibilities and their ratios analytically up to next-to-leading order (NLO) of the HPE. We have shown that the ratio of fourth- to second-order susceptibilities becomes unity in the deconfined phase and nine in the confined phase at leading order. It is also shown that the order of NLO terms, $\kappa^{N_t}$, is significantly high due to the strong cancellation in the ratio.

We have also investigated the properties of thermal excitations in both the deconfined and confined phases. Comparing the grand potential with that of the lattice Boltzmann gas, we have shown that the fermionic excitations in the deconfined phase are the quarks having the excitation energy~\eqref{eq:E_quark2}.
In the confined phase, on the other hand, excitations are dominated by baryonic states carrying the quark charge $\pm3$. We obtained a relation that their excitation energies and degeneracies must satisfy, and decomposed it into flavor multiplets. We have also argued that the baryonic states in the strong-coupling limit are given by a simple form. 

The expectation values of the loop operators that appeared in these results are calculable in Monte Carlo simulations of SU(3) Yang-Mills theory. It is interesting to measure them numerically for further investigation of the fermionic excitations in heavy-quark QCD both in the confined and deconfined phases. The extension of the analysis to higher-order terms in the HPE is another interesting future project. This would allow us to study fermionic excitations around the critical point in heavy-quark QCD~\cite{Cuteri:2020yke,Ashikawa:2024njc}.

\section*{Acknowledgment}
We thank Frithjof Karsch and Hideo Suganuma for valuable discussions.
This work was supported in part by JSPS KAKENHI Grant Numbers JP22K03619, JP23H04507, 24K07049 and 25KJ1618, and ISHIZUE 2025 of Kyoto University.
K.R. and C.S. acknowledge the support by the Polish National Science Centre (NCN) under OPUS Grant No. 2022/45/B/ST2/01527. K.R. also acknowledges the support of the Polish Ministry of Science and Higher Education. 
The work of C.S. was supported in part by the WPI program “Sustainability with Knotted Chiral Meta Matter (WPISKCM2)” at Hiroshima University.
\appendix

\section{Cumulant expansion}
\label{app:cumulant_exp}

In this appendix, after introducing the cumulant expansion in Sec.~\ref{app:exp}, we derive Eq.~\eqref{eq:Omega_q}.

\subsection{Cumulant and cumulant expansion}
\label{app:exp}

Let $P(x)$ be a probability distribution function of a continuous stochastic variable $x$, which satisfies the normalization condition $\int dx P(x)=1$. Cumulants of $x$ are defined as 
\begin{align}
    \langle x^m \rangle_{\rm c} 
    \equiv \frac{\partial^m}{\partial\theta^m} \ln \langle e^{\theta x}\rangle_P \Big|_{\theta=0} ,
    \label{eq:xcum}
\end{align}
where throughout this appendix $\langle \cdots \rangle_P$ represents the expectation value with respect to $P(x)$
\begin{align}
    \langle f(x) \rangle_P \equiv \int dx f(x) P(x) .
    \label{eq:<f(x)>}
\end{align}
Equation~\eqref{eq:xcum} implies that 
\begin{align}
    \ln \langle e^{\theta x}\rangle_P 
    = \sum_{m=1}^\infty \frac{\langle x^m \rangle_{\rm c}}{m!}\theta^m.
\end{align}
Substituting $\theta=1$, we obtain the cumulant expansion
\begin{align}
    \ln \langle e^x\rangle 
    = \sum_{m=1}^\infty \frac{\langle x^m \rangle_{\rm c}}{m!} ,
    \label{eq:cum_exp1}
\end{align}
provided the convergence of the expansion at $\theta=1$.

For the probability distribution $P(x_1,x_2,\cdots,x_N)$ for $N$ multiple stochastic variables $x_1,\cdots,x_N$, the (mixed) cumulants are defined as 
\begin{align}
    \langle x_1^{m_1}\cdots x_N^{m_N}\rangle_{\rm c} 
    = \frac{\partial^{m_1}}{\partial\theta_1^{m_1}}\cdots \frac{\partial^{m_N}}{\partial\theta_N^{m_N}} \ln \Big\langle \exp\sum_{k=1}^N \theta_k x_k\Big\rangle_P \Big|_{\vec\theta=0} ,
    \label{eq:xcummult}    
\end{align}
where $\vec\theta=(\theta_1, \cdots, \theta_N)$.
From Eq.~\eqref{eq:xcummult}, the cumulant expansion for this case is given by
\begin{align}
    &\ln \Big\langle \exp\sum_k \theta_k x_k\Big\rangle_P
    \notag \\
    &= \sum_{m_1,m_2,\cdots,m_N} \frac{ \langle (\theta_1 x_1)^{m_1}\cdots (\theta_N x_N)^{m_N}\rangle_{\rm c}}{m_1!\cdots m_N!}
    \notag \\
    &= \bigg\langle \prod_k \sum_{m_k} \frac{(\theta_k x_k)^{m_k}}{m_k!} \bigg\rangle_{\rm c} .
    \label{eq:cum_exp2}
\end{align}

\subsection{Derivation of Eq.~\eqref{eq:Omega_q}}

Now, let us apply the cumulant expansion to Eq.~\eqref{eq:ZL}. We define the cumulants of $\Lambda_k$ as
\begin{align}
    &\left \langle 
         \Lambda_{k_{1}}^{m_{k_{1}}}  
        \cdots
         \Lambda_{k_{n}}^{m_{k_{n}}}
    \right \rangle_{\scriptsize{\rm c}} \nonumber \\
    &= 
    \frac{\partial^{m_{k_{1}}}}{\partial J_{k_{1}}^{m_{k_{1}}} }
    \cdots
    \frac{\partial^{m_{k_{n}}}}{\partial J_{k_{n}}^{m_{k_{n}}} }
    \ln \left\langle
    \exp 
    \left[
        \sum_{k=- \infty}^{\infty}
        J_{k} \Lambda_{k}
    \right] \right\rangle \Bigg|_{\vec J=0},
    \label{eq:Lcum}
\end{align}
where the bracket $\langle\cdots\rangle$ is defined in Eq.~\eqref{eq:<>}. Their explicit forms up to the third order are shown in Eqs.~\eqref{eq:cumulant1}--\eqref{eq:cumulant3}.
Using Eq.~\eqref{eq:Lcum}, the same procedure as the derivation of Eq.~\eqref{eq:cum_exp2} with assignments $\theta_k$ and $x_k$ with $e^{k\hat\mu_{\rm q}}$ and $\Lambda_k$, respectively, leads to Eq.~\eqref{eq:Omega_q}.

\bibliographystyle{apsrev4-2}
\bibliography{refs}

\end{document}